\documentclass[notitlepage,nofootinbib,preprintnumbers,amssymb,superscriptaddress]{revtex4-1}

\usepackage{amsfonts,amssymb,mathtools,graphicx,color,bm}
\definecolor{ultramarine}{rgb}{0.07, 0.04, 0.56}
\definecolor{cadmiumgreen}{rgb}{0.0, 0.42, 0.24}
\definecolor{indigo(dye)}{rgb}{0.0, 0.25, 0.42}
\usepackage[linktocpage=true,breaklinks]{hyperref}
\hypersetup{
colorlinks=true,
citecolor=ultramarine,
linkcolor=cadmiumgreen,
urlcolor=indigo(dye),
}

\usepackage{braket}

\newcommand{\fr}[2]{\frac{#1}{#2}}
\newcommand{\pa}{\partial}
\newcommand{\ti}{\tilde}
\newcommand{\na}{\nabla}
\newcommand{\braa}[1]{\left( #1 \right)}  
\newcommand{\brab}[1]{\left[ #1 \right]}  
\newcommand{\brac}[1]{\left\{ #1 \right\}}  
\newcommand{\be}{\begin{equation}}  
\newcommand{\ee}{\end{equation}}
\newcommand{\bem}{\begin{bmatrix}}
\newcommand{\eem}{\end{bmatrix}}
\newcommand{\Mpl}{M_{\rm Pl}}
\newcommand{\da}{\dagger}

\newcommand{\ep}{\epsilon}

\newcommand{\la}{\lambda}
\newcommand{\si}{\sigma}

\newcommand{\mn}{{\mu \nu}}

\newcommand{\mO}{\mathcal{O}}

\newcommand{\ve}[1]{{\rm{\mathbf #1}}}

\begin{document}

\preprint{KOBE-COSMO-19-14, MAD-TH-19-08}

\title{Effective Field Theory of Anisotropic Inflation and Beyond}

\author{Jinn-Ouk Gong}
\affiliation{Korea Astronomy and Space Science Institute, Daejeon 34055, Korea}

\author{Toshifumi Noumi}
\affiliation{Department of Physics, Kobe University, Kobe 657-8501, Japan}

\author{Gary Shiu}
\affiliation{Department of Physics, University of Wisconsin-Madison, Madison, WI 53706, USA}

\author{Jiro Soda}
\affiliation{Department of Physics, Kobe University, Kobe 657-8501, Japan}

\author{Kazufumi Takahashi}
\affiliation{Department of Physics, Kobe University, Kobe 657-8501, Japan}

\author{Masahide Yamaguchi}
\affiliation{Department of Physics, Tokyo Institute of Technology, Tokyo 152-8551, Japan}

\begin{abstract}

We develop an effective-field-theory (EFT) framework for inflation with various symmetry breaking pattern. As a prototype, we formulate anisotropic inflation from the perspective of EFT and construct an effective action of the Nambu-Goldstone bosons for the broken time translation and rotation symmetries. We also calculate the statistical anisotropy in the scalar two-point correlation function for concise examples of the effective action.

\end{abstract}

\maketitle


\section{Introduction}
\label{sec:intro}

Cosmic inflation, which accounts for the origin of the large-scale structure of the universe, is strongly supported by observations of cosmic microwave background (CMB)~\cite{Bennett:2012zja, Hinshaw:2012aka, Ade:2013sjv, Planck:2013jfk,Komatsu:2010fb}. As is often referred to as the cosmic no-hair conjecture~\cite{Wald:1983ky}, any classical memory is washed out by the exponential de~Sitter expansion of the universe, hence primordial quantum fluctuations are responsible for the large-scale structure. Remarkably, the nature of the primordial fluctuations is understood by symmetries in inflation which can be summarized as follows:

\begin{itemize}

\item {\bf Spatial translation and rotation symmetry}\\
The cosmic no-hair conjecture states that the universe will be homogenized and isotropized in a few Hubble expansion times once the inflationary epoch starts. The universe then enjoys the spatial translation and rotation symmetries.
In particular, the universe is led to de~Sitter space if the vacuum energy is dominant.

\item {\bf de~Sitter dilatation symmetry}\\
The metric for de~Sitter space reads 
\begin{eqnarray}
 ds^2 = -dt^2 + e^{2 Ht} ( dx^2 + dy^2 + dz^2) \,,
\end{eqnarray}
where the Hubble parameter $H$ is constant. It is easy to find the dilatation symmetry:
\begin{equation}
t \rightarrow t+ \epsilon \,,\quad  x^i \rightarrow e^{-H \epsilon} x^i \,, \quad x^i =\{x,y,z\} \,,
\end{equation}
where $\epsilon$ being a constant parameter. Also, de~Sitter space accommodates the special conformal symmetry accompanied by an appropriate transformation of the time coordinate.

\item {\bf Shift symmetry of the inflaton field}\\ In addition to the
above spacetime symmetries, there exists a symmetry in the field space
for a single-field inflation with a canonical kinetic term. In order to
have slow-roll inflation, we need a sufficiently flat potential, hence
the inflaton field~$\phi (x)$ is required to enjoy a shift symmetry,
$\phi (x) \rightarrow \phi (x) + {\rm const}$.
\end{itemize}

The above symmetries can be promoted to statistical symmetries which
determine the nature of primordial fluctuations. In general, we need
$n$-point correlation functions to characterize the statistical nature
of primordial fluctuations. For simplicity, we assume a single-field
inflation with a standard kinetic term, which implies the shift symmetry
in the field space. Then, the shift symmetry forbids nonlinearity, and
hence the statistics of fluctuations should be Gaussian.\footnote{
Strictly speaking, the shift symmetry allows not only the canonical
kinetic term but also more generic forms of kinetic terms, whose
nonlinearity can lead to non-Gaussian fluctuations.} Thus, two-point
functions carry all the statistical information.

Let us next implement the de~Sitter space symmetry.
For an illustration, we consider the curvature perturbation~$\zeta$.
Because of the spatial translation symmetry, we can work in the Fourier space.
Then, we have the power spectrum:
\begin{equation}
\braket{\zeta ({ \bf k} )  \, \zeta ({ \bf k}' )} = (2\pi)^3\delta^{(3)}({ \bf k}+{ \bf k}')\frac{2\pi^2}{k^3}P ({ \bf k}) \,,
\end{equation}
where ${\bf k}$ and ${\bf k}'$ are the wavenumber vectors and $k=|{\bf k}|$. 
The delta function stems from  ``momentum'' conservation. 
Note that the dipole-type anisotropy is prohibited because
\begin{equation}
\braket{\zeta ({ \bf k} )  \, \zeta (-{ \bf k} )}
=\braket{\zeta (-{ \bf k} )  \, \zeta ({ \bf k} )}
\quad
\leftrightarrow
\quad
P ( {\bf k}) =  P (- {\bf k}) \,,
\end{equation}
where we have assumed that the modes are outside of the horizon so that $\zeta ({ \bf k} )$ and $\zeta (-{ \bf k} )$ commute with each other. 
The rotation symmetry further constrains the power spectrum as 
\begin{equation}
\braket{\zeta ({ \bf k} )  \, \zeta ({ \bf k}' ) } = (2\pi)^3\delta^{(3)}({ \bf k}+{ \bf k}')\frac{2\pi^2}{k^3}P (k=|{\bf k}|) \,.
\end{equation}
Namely, the direction dependence of the power spectrum is forbidden by the rotation symmetry. 
Finally, the de~Sitter dilatation symmetry implies a scale-invariant power spectrum,
\begin{equation}
P(k) = {\rm const}\,,
\end{equation}
because the curvature perturbation is conserved on large scales. 
Thus, the statistically homogeneous, isotropic, and scale-invariant Gaussian primordial curvature fluctuations are a consequence of the de~Sitter symmetry.
These predictions are robust and universal in conventional single-field inflationary scenarios.
In fact, the above predictions have been almost confirmed by CMB observations~\cite{Komatsu:2010fb}.

However, when we look at the fine structure of the CMB data, we need to elaborate on the theoretical details.
In fact, the cosmic expansion is not exactly de~Sitter but quasi-de~Sitter. 
Namely, there exists violation of the de~Sitter dilatation symmetry due to the time-dependence of the inflaton background $\braket{\dot{\phi}(x)} =\dot{\bar{\phi}}(t)\neq0$, which leads to a slight tilt of the power spectrum. 
Here, the bar denotes an expectation value. 
As the deviation from the de~Sitter expansion can be characterized by the slow-roll parameter, the tilt should be of the order of the slow-roll parameter. 
The deviation from the de~Sitter expansion also allows the model to have nonlinearity of the order of the slow-roll parameter, namely, non-Gaussianity~\cite{Maldacena:2002vr}. 
Moreover, if we consider multi-field inflation or a non-standard kinetic term, we can have nonlinearity, which leads to a sizable non-Gaussianity~\cite{Bernardeau:2002jy, Chen:2006nt}. 
Remarkably, these fine structures of the primordial fluctuations have been observationally tested.
The spectral tilt has been well confirmed~\cite{Komatsu:2010fb} and non-Gaussianity has been strongly constrained~\cite{Ade:2015ava}.

Along this line of thought, it is natural to expect violation of the spatial rotation symmetry, which would lead to the statistical anisotropy~\cite{ArmendarizPicon:2005jh,ArmendarizPicon:2007nr}. 
Intriguingly, WMAP provided some hints of the statistical anisotropy~\cite{Eriksen:2003db,Schwarz:2004gk,deOliveira-Costa:2003utu,Bennett:2010jb}.
Hence, there appeared serious studies of quadrupole anisotropy~\cite{Ackerman:2007nb,Pullen:2007tu,ArmendarizPicon:2008yr}.
Indeed, significant quadratic anisotropy in the power spectrum was reported in~\cite{Groeneboom:2008fz,Groeneboom:2009cb}.
Motivated by these studies, many theoretical mechanisms for realizing the statistical anisotropy have been proposed~\cite{Gumrukcuoglu:2006xj,Yokoyama:2008xw,Karciauskas:2008bc,Tahara:2018orv}. 
These attempts were challenges to the cosmic no-hair conjecture.  
A stable anisotropic inflation model was found in the context of supergravity~\cite{Watanabe:2009ct}, which gives rise to a clear counterexample to the cosmic no-hair conjecture~\cite{Kanno:2010nr}. 
There can be various extension of anisotropic inflation~\cite{Soda:2012zm,Maleknejad:2012fw}.
In these models, the statistical anisotropy appears in the form~\cite{Dulaney:2010sq,Gumrukcuoglu:2010yc,Watanabe:2010fh}
\begin{eqnarray}
P({\bf k}) = P(k) \left[ 1+ g_* \sin^2 \theta_\ve{k} \right] \,,
\end{eqnarray}
where $P(k)$ is the isotropic part of the power spectrum $P({\bf k})$ and $\theta_\ve{k}$ is the angle between the preferred direction ${\bf n}$ and the wavenumber vector ${\bf k}$ of fluctuations.
 Here, $g_*$ characterizes the magnitude of the statistical anisotropy.
In fact, the way to make a preferred direction is not unique. 
For example, anisotropic inflation with two-form filed is 
possible~\cite{Watanabe:2010fh,Ohashi:2013mka,Ito:2015sxj}.
The preferred direction ${\bf n}$ specified by a two-form field can be translated to the linear inhomogeneity of the scalar field through the dual transformation. 
This is nothing but the anisotropy in
solid inflation~\cite{Endlich:2012pz,Bartolo:2013msa}. 
A useful scheme for studying such models in a unified manner is the effective field theory~(EFT) approach~\cite{Cheung:2007st}. 
A direct application of EFT method to anisotropic inflation has been first performed in \cite{Abolhasani:2015cve,Rostami:2017wiy}.
As the theoretical investigation progressed, it turned out that 
the observed quadrupole anisotropy can be attributed to the anisotropy of the beam~\cite{Hanson:2010gu}.
Now, we only have an upper bound $g_* = 0.002\pm 0.016$ (68$\%$ CL)~\cite{Kim:2013gka}.\footnote{A constraint on the background anisotropy was given in \cite{Bunn:1996ut}, which is much looser than the one on the statistical anisotropy. For a more detailed discussion, see \S \ref{ssec:nondynamicalh0}.}
Of course, this does not mean the statistical anisotropy does not exist. 
Rather, the anisotropic inflation opens up a new direction of research. Therefore, it is worthy of study in detail.

So far, we have not discussed the violation of spatial translation symmetry.
However, since the Planck observations give a hint of the dipole anisotropy~\cite{Ade:2015hxq,Schwarz:2015cma}, 
we have a good motivation to consider this possibility.
We can also consider the violation of Lorentz boost symmetry\footnote{As was shown in \cite{ArmendarizPicon:2010mz}, the effective action for the case with broken boost symmetry has a close connection to the Einstein-aether model~\cite{Jacobson:2000xp}.}~\cite{ArmendarizPicon:2010mz,Delacretaz:2015edn} or spatial diffeomorphisms~\cite{Cannone:2014uqa,Bartolo:2015qvr}. 
Apparently, we need a broader framework to incorporate these possibilities.
The EFT approach~\cite{Hidaka:2014fra} based on the spacetime symmetry viewpoint should be useful for this purpose. 
Indeed, the effective theory can accommodate more exotic models such as anisotropic k-inflation
where the kinetic term is non-standard~\cite{Ohashi:2013pca}.
The aim of this paper is to develop an EFT formalism of anisotropic inflation from the perspective of general spacetime symmetry breaking.  The formalism allows us
to go beyond anisotropic inflation.

In this article, we illustrate our idea by formulating the EFT of
anisotropic inflation. 
Following \cite{Hidaka:2014fra}, we introduce
general local Lorentz frames to describe spontaneous breaking of spacetime symmetries and construct EFTs in the unitary gauge. Then, we introduce three Nambu-Goldstone~(NG) bosons~(which will be called $\eta_0$, $\eta_1$, and $\eta_2$) for the broken spatial rotation symmetry on top of the one~(called $\pi$) for the broken time translation symmetry.
It should be noted that one of the NG modes corresponding to the broken rotation symmetry (i.e., $\eta_0$) is nondynamical for the conventional model of anisotropic inflation having the standard Maxwell kinetic term. 
Our EFT framework covers not only such models with nondynamical $\eta_0$ but also those with dynamical $\eta_0$, which are not covered by the framework of \cite{Abolhasani:2015cve,Rostami:2017wiy}.
We shall discuss differences between this work and the EFT approach in \cite{Abolhasani:2015cve,Rostami:2017wiy} more in detail in \S \ref{sec:conclusions}.

The rest of this article is organized as follows.
In \S \ref{sec:example}, we clarify the symmetry breaking patterns and specify the associated NG modes for a concrete model of anisotropic inflation.
We also explain the unitary gauge conditions, under which all the NG modes are eaten by the vierbein and the gauge field.
In \S \ref{sec:EFT}, we construct the effective action in the unitary gauge and discuss the tadpole cancellation conditions.
In \S \ref{sec:aniso}, we compute the scalar two-point correlation function based on the so-obtained EFT framework.
Finally, we draw our conclusions in \S \ref{sec:conclusions}.


\section{Symmetry structures in a concrete model}
\label{sec:example}

As was shown in  \cite{Hidaka:2014fra}, each broken local symmetry has one associated dynamical mode, which gives a correct identification of NG bosons with broken symmetries (see Appendix~\ref{App:review} for a brief review). In this section, we reformulate an anisotropic inflationary model based on such a local spacetime symmetry point of view and 
give a flavor of an EFT framework for anisotropic inflation and
beyond.

Let us consider the following conventional action for a neutral scalar~$\phi$ and an Abelian gauge field~$A_\mu$:
\begin{align}
S=\int d^4x\sqrt{-g}\left[
\frac{\Mpl^2}{2}R
-\frac{1}{2}(\partial_\mu\phi)^2-V(\phi)
-\frac{f(\phi)^2}{4}F_{\mu\nu}F^{\mu\nu}
\right]\,, 
\label{simplest}
\end{align}
where $\Mpl$ is the reduced Planck mass
and $F_{\mu\nu}\coloneqq \partial_\mu A_\nu-\partial_\nu A_\mu$ is the field strength of the gauge field.
If we introduce the vierbein~$e^m_\mu$ and rewrite the gauge field as
\begin{equation}
A_\mu=e_\mu^mA_m\,,
\end{equation}
we can write down a local-Lorentz-invariant action.
Here and in what follows, we use Greek letters for the curved-spacetime indices ($\mu,\nu,\cdots=0,1,2,3$) and Latin letters for the (local-)Minkowski indices ($m,n,\cdots={\bar0},{\bar1},{\bar2},{\bar3}$).
Let us now assume the spatially homogeneous anisotropic background:
\begin{align}
\label{eq:BGmetric}
ds^2=-dt^2+a(t)^2\left[e^{2\sigma(t)}\left(dx^2+dy^2\right)+e^{-4\sigma(t)}dz^2\right]\,.
\end{align}
Here, without loss of generality we have given the anisotropy along the $z$-direction.
Then, the background quantities are
\begin{align}
\label{background_real}
\phi(x)=\bar\phi(t)\,,
\quad
A_\mu(x)=e_\mu^{\bar3}\bar v(t)\,,
\quad
e_\mu^m=\text{diag}\left(1,ae^{\sigma},ae^{\sigma},ae^{-2\sigma}\right)\,.
\end{align}
The independent components of the Einstein equation are given by
\begin{align}
3\Mpl^2(H^2-\dot{\si}^2)
&=
\frac{1}{2}\dot{\bar{\phi}}^2
+V(\bar{\phi})+\fr{f(\bar{\phi})^2}{2}\braa{\dot{\bar{v}}+H\bar{v}-2\dot{\si}\bar{v}}^2\,, 
\label{bgEinstein00}
\\
\Mpl^2(3H^2+2\dot{H}-3H\dot{\si}-\ddot{\si}+3\dot{\si}^2)
&=
-\frac{1}{2}\dot{\bar{\phi}}^2+V(\bar{\phi})-\fr{f(\bar{\phi})^2}{2}\braa{\dot{\bar{v}}+H\bar{v}-2\dot{\si}\bar{v}}^2\,, \label{bgEinstein11}
\\
\Mpl^2(3H^2+2\dot{H}+6H\dot{\si}+2\ddot{\si}+3\dot{\si}^2)&=
-\frac{1}{2}\dot{\bar{\phi}}^2+V(\bar{\phi})+\fr{f(\bar{\phi})^2}{2}\braa{\dot{\bar{v}}+H\bar{v}-2\dot{\si}\bar{v}}^2\,. \label{bgEinstein33}
\end{align}
The equation of motion for the vector field is
\begin{equation}
\pa_t\brab{f(\bar{\phi})^2a^2e^{2\si}\braa{\dot{\bar{v}}+H\bar{v}-2\dot{\si}\bar{v}}}=0\,, 
\label{bgEOMvec}
\end{equation}
while the one for the scalar field is automatically satisfied for the solution of \eqref{bgEinstein00}--\eqref{bgEOMvec}.

We now discuss perturbations around the background, emphasizing on its symmetry breaking pattern and the associated NG modes.
Since the background~\eqref{background_real} breaks the time translation symmetry and the rotation symmetry associated with the $z$-direction, we have two types of NG modes. 
To identify those degrees of freedom, it is convenient to decompose $\phi$ and $A_\mu$ as
\begin{equation}
\phi(x)=\bar{\phi}\braa{t+\pi(x)}\,,
\quad
A_\mu(x)=e^m_\mu\Lambda_m{}^{\bar3}\bar{v}\braa{t+\chi(x)}\,.
\end{equation}
Here, one linear combination of $\pi$ and $\chi$ is the NG mode for the
broken time translation. Since the $U(1)$ symmetry is not
spontaneously broken and hence the would-be associated NG mode is not eaten by a
gauge field, the orthogonal linear combination of $\pi$ and $\chi$ is not the
NG mode but a gauge degree of freedom.
Three NG modes, $\eta_{\widehat m}$ (${\widehat m}=0,1,2$), for the broken rotation symmetry are encoded in $\Lambda_m{}^n\in SO(3,1)$ as
    \begin{equation}
    \Lambda_{\widehat m}{}^{\bar3}=\eta_{\widehat m}+\mO(\eta^3)\,,
    \quad
    \Lambda_{\bar3}{}^{\bar3}=1+\fr{1}{2}\braa{\eta_0^2-\eta_1^2-\eta_2^2}+\mO(\eta^4)\,.
    \end{equation}
The NG bosons~$\eta_{\widehat{m}}$ transform as a Lorentz 3-vector under the unbroken $SO(2,1)$ transformations and as scalars under diffeomorphisms.
A caveat should be added here.
One could attempt to describe fluctuations of the gauge field $A_\mu$, e.g., as an ``NG boson'' for spatial diffeomorphism, but it turns out not so useful as we now explain. Suppose that the third component of the vector field~$A_\mu$ has a time-dependent background as
\begin{align}
A_\mu=\delta_\mu^3\bar{V}(t) 
\,,
\end{align}
where $\bar{V}(t)\coloneqq ae^{-2\sigma}\bar{v}$. Analogously to the NG boson for the spatial diffeomorphism in the third direction, one would introduce a scalar field $\psi$ as
\begin{align}
A_\mu= \bar{V}(t)\left(\delta_\mu^3+\partial_\mu\psi\right)\,.
\end{align}
The fluctuation of the vector field~$\delta A_\mu=A_\mu-\delta_\mu^3\bar{V}(t)$ is then related to $\psi$ as
\begin{align}
\delta A_\mu=\bar{V}(t)\partial_\mu\psi\,, 
\end{align}
through a derivative. First of all, we cannot describe the four components of the vector fluctuations in terms of a single scalar $\psi$.   Also, the constant mode of $\psi$ does not generate any fluctuation of $\delta A_\mu$, essentially because the translation symmetry along the third direction is not broken. This is in sharp contrast to the standard NG bosons and indicates a wrong identification of NG bosons. On the other hand, if we introduce NG bosons for local Lorentz symmetries as above, the NG bosons provide appropriate degrees of freedom and also they are identified with vector fluctuations without involving derivatives. See, e.g.,~\cite{Hidaka:2014fra} for a more detailed discussion.

We next introduce the action in the unitary gauge, where all the 
NG modes are eaten by the gauge field and the vierbein. 
We fix the gauge degrees of freedom for the time diffeomorphism and three of the six local Lorentz symmetries by $\pi=0$ and $\eta_{\widehat m}=0$, respectively.
Regarding $\chi$, one can use the $U(1)$ gauge degree of freedom to set $\chi=0$.
In this gauge, the action reduces to the form:
    \begin{align}
    S&=S_{\rm EH}+S_\phi+S_A\,, 
    \\
    \quad
    S_{\rm EH}&=\fr{\Mpl^2}{2}\int d^4x\sqrt{-g}R\,, 
    \\
    \quad
    S_\phi&=\int d^4x\sqrt{-g}
    \brab{ - \frac{1}{2}\dot{\bar{\phi}}^2g^{00} - V(\bar{\phi}) }\,, 
    \\
    S_A&=\int d^4x\sqrt{-g}f(\bar\phi)^2\brab{-\fr{\dot{\bar{v}}^2}{2}g^{00}-\frac{\bar{v}^2}{4}\left(\partial_\mu e_\nu^{\bar3}-\partial_\nu e_\mu^{\bar3}\right)^2+\fr{\dot{\bar{v}}^2}{2}\left(e^{0\bar3}\right)^2+\bar{v}\dot{\bar{v}}\delta_\mu^0\left(e^{\nu\bar3}\nabla_\nu e^{\mu\bar3}\right)}\,.\label{S_A}
    \end{align}
Here, the first term in \eqref{S_A} breaks the time-diffeomorphism symmetry only, while the rest terms with the vierbein break the local Lorentz invariance (of which the last two terms containing $\delta^0_\mu$ break also the time diffeomorphism).
In this gauge, the metric accommodates three propagating degrees of freedom (two helicity modes of a massless graviton and the inflaton eaten by the metric) and the vierbein does two (two helicity modes of photon eaten by the vierbein).
Note that the residual gauge symmetries in our gauge choice are spatial diffeomorphism symmetries and three local Lorentz symmetries along the $\bar0,\bar1,\bar2$ directions, which are manifest in the above action.
Note also that here and in what follows $e^{\bar3}_\mu$ should be understood as $e^m_\mu\delta^{\bar3}_m$, where the local-Lorentz indices are contracted. 
Hence, we treat $e^{\bar3}_\mu$ as an ordinary four-vector and define its covariant derivative as $\na_\mu e^{\bar3}_\nu\coloneqq \pa_\mu e^{\bar3}_\nu-\Gamma^\la_\mn e^{\bar3}_\la$, where $\Gamma^\la_\mn$ is the Levi-Civita connection with respect to $g_\mn$.

Instead of the derivative of the vierbein itself, it is useful to introduce the quantity
    \begin{equation}
    \label{eq:Dmn}
    D_\mn\coloneqq \na_\mu e^{\bar3}_\nu-e^{\bar3}_\mu n_\nu(H-2\dot{\si})\,,
    \end{equation}
which is covariant under the unbroken symmetries and vanishes on the background.
Here, the timelike unit vector~$n_\mu$ is defined by
    \begin{equation}
    n_\mu\coloneqq -\fr{\delta^0_\mu}{\sqrt{-g^{00}}}\,.
    \end{equation}
With this $D_\mn$ and $\delta g^{00}=g^{00}+1$, the total action can be recast as
    \begin{align}
    S=\int d^4x\sqrt{-g}\biggl\{&
    \!\fr{\Mpl^2}{2}R+(c-\Lambda)-c\delta g^{00}+\la D_\mn e^{\mu{\bar3}}n^\nu 
    \nonumber \\
    &-\fr{f(\bar\phi)^2\bar{v}\dot{\bar{v}}}{8}(H-2\dot{\si})\braa{\delta g^{00}}^2
    -\fr{f(\bar\phi)^2\bar{v}^2}{2}\braa{D_\mn D^\mn-D_\mn D^{\nu\mu}} 
    \nonumber \\
    &+\fr{f(\bar\phi)^2}{2}\brab{\dot{\bar{v}}^2-\bar{v}^2(H-2\dot{\si})^2}\braa{n^\mu e_\mu^{\bar3}}^2
    +\fr{f(\bar\phi)^2\bar{v}\dot{\bar{v}}}{2}\delta g^{00}D_\mn e^{\mu\bar{3}}n^\nu+\mO(\delta^3)
    \biggr\}\,, 
    \label{simpleUGaction}
    \end{align}
with $c$, $\Lambda$, and $\la$ being
    \begin{align}
    c&=\frac{1}{2}\dot{\bar{\phi}}^2
    +\fr{f(\bar\phi)^2\dot{\bar{v}}}{2}\braa{\dot{\bar{v}}+H\bar{v}-2\dot{\si}\bar{v}}\,, 
    \\
    \Lambda&=V(\bar{\phi})-\fr{f(\bar\phi)^2\bar{v}}{2}\braa{H-2\dot{\si}}\braa{\dot{\bar{v}}+H\bar{v}-2\dot{\si}\bar{v}}\,, 
    \\
    \la&=-f(\bar\phi)^2\bar{v}\braa{\dot{\bar{v}}+H\bar{v}-2\dot{\si}\bar{v}}\,.
    \end{align}
Also, $\mathcal{O}(\delta^n)$ denotes $n$th and higher-order terms in fluctuations. In \eqref{simpleUGaction}, we have suppressed terms cubic or higher order in fluctuations.

Thus, we have reformulated the original model of anisotropic inflation~\cite{Watanabe:2009ct}.
In the next section, we construct general EFTs of anisotropic inflation.


\section{Effective action in the unitary gauge}
\label{sec:EFT}

As we illustrated in the previous section, the field contents and the residual gauge symmetries in the unitary gauge for our problem are
\begin{align}
\label{eq:symms}
e_\mu^m\,,
\quad
\text{matters}
\quad
+\quad
\text{spatial diffs}\,,\,
\text{(2+1)-dim local Lorentz}\,,
\end{align}
where the matters are non-NG fields, i.e. the degrees of freedom which cannot be absorbed by gauge transformations.
We then construct the effective action for the NG fields without matters.
The building blocks are any quantities that transform as a tensor under spatial diffeomorphisms and a scalar under $SO(2,1)$ transformations, namely, the metric~$g_\mn$, the timelike unit vector~$n_\mu$, the $\bar{3}$-component of the vierbein~$e^{\bar 3}_\mu$, and their derivatives.
For simplicity, we impose the invariance under the $Z_2$ transformation $e_\mu^{\bar3}\to-e_\mu^{\bar3}$ as well.

Under these assumptions, it is shown in Appendix~\ref{AppA} that the general unitary-gauge action up to the linear order in perturbations takes the form
    \begin{equation}
    \label{general_linear}
    S=\int d^4x\sqrt{-g}\brab{\fr{\Mpl^2}{2}R-\Lambda-c g^{00}+\la D_\mn e^{\mu{\bar3}}n^\nu+\mO(\delta^2)}\,,
    \end{equation}
where $\Lambda$, $c$, and $\lambda$ are scalar functions of time. Also, we took the Einstein frame to fix the coefficient of the Ricci scalar. Note that terms with more derivatives (e.g. those including the extrinsic curvature $K_\mn$) can be converted to higher-order terms up to total derivatives (see~\cite{Cheung:2007st} for details in the case of single-field inflation). For a generic choice of EFT parameters~$(\Lambda,c,\lambda)$, the action~\eqref{general_linear} contains tadpole terms. Requiring that the background satisfies the equation of motion, we remove them by
the following tadpole cancellation conditions: 
\begin{align}
\Lambda
&=M_{\rm Pl}^2
\left(
3H^2
+\dot{H}
-\frac{3}{2}H\dot{\sigma}-\frac{1}{2}\ddot{\sigma}
\right)
+\frac{1}{2}(H-2\dot{\sigma})\lambda\,, \label{tadpole1}
\\
c&=
M_{\rm Pl}^2\left(-\dot{H}+\frac{3}{2}H\dot{\sigma}+\frac{1}{2}\ddot{\sigma}-3\dot{\sigma}^2\right)
+\frac{1}{2}(H-2\dot{\sigma})\lambda\,,
\\
3H\lambda+\dot{\lambda}&=-M_{\rm Pl}^2(9H\dot{\sigma}+3\ddot{\sigma})\,, \label{tadpole3}
\end{align}
which enables us to write down the EFT parameters in terms of background geometry.
For the detail of the derivation, see Appendix~\ref{AppA}.

Let us then incorporate higher-order terms in the perturbations. Since we are interested in the primordial two-point functions which will be discussed in the next section, we focus on the second-order action, even though it is straightforward to generalize the analysis to higher-order terms, e.g., relevant to non-Gaussianities. We also focus on the leading order in derivatives where each NG mode carries at most one derivative. Under these assumptions, the general action in the unitary gauge up to the quadratic order in perturbations can be written as
    \begin{align}
    S=\int d^4x\sqrt{-g}\bigg[&\fr{\Mpl^2}{2}R+(c-\Lambda)-c\delta g^{00}+\la D_\mn e^{\mu{\bar3}}n^\nu \nonumber 
    +\fr{M_2^4}{2}\braa{\delta g^{00}}^2
     \nonumber \\
    &-\fr{m_1^2}{2}D_\mn D^\mn+\fr{m_2^2}{2}D_\mn D^{\nu\mu}+\fr{m_3^4}{2}\braa{n^\mu e_\mu^{\bar3}}^2+\fr{m_4^3}{2}\delta g^{00}D_\mn e^{\mu{\bar3}}n^\nu 
    \nonumber \\
    &+\fr{m_5^2}{2}\braa{D^\mu_\mu}^2+\fr{m_6^2}{2}\braa{D_\mn e^{\mu\bar{3}}}^2+\fr{m_7^3}{2}D^\mu_\mu n^\nu e_\nu^{\bar3}+\fr{m_8^2}{2}\braa{D_\mn n^\mu}^2 
    \nonumber \\
    &-\fr{m_9^2}{2}\braa{D_\mn n^\nu}^2+\fr{m_{10}^2}{2}D_\mn D^{\nu\la}n^\mu n_\la+\mO(\delta^3)\biggr]\,.
     \label{SeffUG}
    \end{align}
Here, we have neglected terms with the curvature tensor as before since they merely amount to a redefinition of the existing coefficients  up to total derivative.
For instance, terms like $R_{{\bar 3}\mu}{}^{{\bar 3}\mu}$ can be absorbed into \eqref{SeffUG} by using $R_{\mn}e^{\nu{\bar 3}}=[\na_\mu,\na_\nu]e^{\nu{\bar 3}}$ and integration by parts.
The lowest-order EFT parameters~$(\Lambda,c,\lambda)$ follow from the tadpole cancellation conditions~\eqref{tadpole1}--\eqref{tadpole3}.
Note in passing that we have omitted higher derivative terms with $\delta K_{\mu\nu}$ or $\delta R_{\mu\nu\la\si}$ in \eqref{SeffUG}.
When taking into account such terms, one must be careful as the extrinsic curvature and the Riemann tensor now contain anisotropic contributions. Their background values, $\bar{K}^\mu_\nu$ and $\bar{R}^{\mu\nu}{}_{\lambda\sigma}$, are written as follows\footnote{
Note that the background value of the Weyl tensor is calculated as
    \be
    \bar{C}^\mn{}_{\la\si}=-2\braa{H\dot{\si}-2\dot{\si}^2+\ddot{\si}}
    \braa{h^{\mu}_{[\la}h^{\nu}_{\si]}+h^{[\mu}_{[\la}n^{\nu]}n_{\si]}-3h^{[\mu}_{[\la}e^{\nu]\bar{3}}e_{\si]}^{\bar{3}}-3n^{[\mu}n_{[\la}e^{\nu]\bar{3}}e_{\si]}^{\bar{3}}}\,, \nonumber
    \ee
which vanishes in the isotropic case. Its fluctuation, $\delta C^{\mu\nu}{}_{\lambda\sigma}=\delta C^{\mu\nu}{}_{\lambda\sigma}-\bar{C}^{\mu\nu}{}_{\lambda\sigma}$, will also be useful in the EFT context because it describes propagating graviton modes which cannot be eliminated by field redefinition or equivalently by using the Einstein equation.}:
    \begin{align}
    \bar{K}^\mu_\nu&=\braa{H+\dot{\si}}h^\mu_\nu-3\dot{\si}e^{\mu\bar{3}}e_{\nu}^{\bar{3}}\,, \\
    \bar{R}^\mn{}_{\la\si}&=
    2\braa{H+\dot{\si}}^2h^{\mu}_{[\la}h^{\nu}_{\si]}
    -4\braa{H+\dot{H}+2H\dot{\si}+\dot{\si}^2+\ddot{\si}}h^{[\mu}_{[\la}n^{\nu]}n_{\si]} \nonumber \\
    &\quad 
    -12\braa{H\dot{\si}+\dot{\si}^2}h^{[\mu}_{[\la}e^{\nu]\bar{3}}e_{\si]}^{\bar{3}}
    +12\braa{2H\dot{\si}-\dot{\si}^2+\ddot{\si}}n^{[\mu}n_{[\la}e^{\nu]\bar{3}}e_{\si]}^{\bar{3}}\,,
    \end{align}
where $h_\mn\coloneqq g_\mn+n_\mu n_\nu$ is the induced metric on constant time slices.
Their fluctuations in the unitary gauge are then defined as $\delta K^\mu_\nu=K^\mu_\nu-\bar{K}^\mu_\nu$ and $\delta R^{\mu\nu}{}_{\lambda\sigma}=\delta R^{\mu\nu}{}_{\lambda\sigma}-\bar{R}^{\mu\nu}{}_{\lambda\sigma}$ in a covariant manner to respect the residual gauge symmetries in the unitary gauge.

In this section, we have succeeded in formulating EFTs with broken time translation and rotation symmetries.
It should be stressed that the procedure performed in this section can be generalized to other inflationary models with various symmetry breaking patterns.


\section{Statistical anisotropy}
\label{sec:aniso}

In this section, we study the statistical anisotropy based on the effective action obtained in the previous section.
To this end, we reintroduce the NG bosons~$\pi$ and $\eta_{\widehat m}$ via the St{\"u}ckelberg trick, and then construct the quadratic action for the NG bosons under the decoupling limit.
We first discuss the conventional model considered in \S \ref{sec:example}, where $\eta_0$ is nondynamical.
Then, we investigate the case with dynamical $\eta_0$, taking a simple example.

\subsection{Nondynamical $\eta_0$}
\label{ssec:nondynamicalh0}

Let us focus on the simple model discussed in \S \ref{sec:example}, for which we have the following effective action in the unitary gauge:
    \begin{align}
    S=\int d^4x\sqrt{-g}\biggl[&
    \fr{\Mpl^2}{2}R+(c-\Lambda)-c\delta g^{00}+\la D_\mn e^{\mu{\bar3}}n^\nu+\fr{M_2^4}{2}\braa{\delta g^{00}}^2
    -\fr{m_1^2}{2}D_\mn D^\mn+\fr{m_2^2}{2}D_\mn D^{\nu\mu} 
    \nonumber \\
    &+\fr{m_3^4}{2}\braa{n^\mu e_\mu^{\bar3}}^2
    +\fr{m_4^3}{2}\delta g^{00}D_\mn e^{\mu\bar{3}}n^\nu+\mO(\delta^3)
    \biggr]\,. \label{SeffUGsimple}
    \end{align}
Here, the parameters are written as
    \begin{align}
    c&=\frac{1}{2}\dot{\bar{\phi}}^2+\fr{f(\bar\phi)^2\dot{\bar{v}}}{2}\braa{\dot{\bar{v}}+H\bar{v}-2\dot{\si}\bar{v}}\,, 
\label{eq:parametersfirst}
    \\
    \Lambda&=V(\bar{\phi})-\fr{f(\bar\phi)^2\bar{v}}{2}\braa{H-2\dot{\si}}\braa{\dot{\bar{v}}+H\bar{v}-2\dot{\si}\bar{v}}\,, 
    \\
    \la&=-f(\bar\phi)^2\bar{v}\braa{\dot{\bar{v}}+H\bar{v}-2\dot{\si}\bar{v}}\,, 
    \\
    M_2^4&=-\fr{f(\bar\phi)^2\bar{v}\dot{\bar{v}}}{4}(H-2\dot{\si})\,, 
    \\
    m_1^2=m_2^2&=f(\bar\phi)^2\bar{v}^2\,, 
    \\
    m_3^4&=f(\bar\phi)^2\brab{\dot{\bar{v}}^2-\bar{v}^2(H-2\dot{\si})^2}\,, 
    \\
    m_4^3&=f(\bar\phi)^2\bar{v}\dot{\bar{v}}\,.
\label{eq:parameterslast}
    \end{align}
Given the fact that the observed anisotropy is tiny, one can use the slow-roll and small-anisotropy approximation~\eqref{approx} to yield
    \begin{align}
    &c=\ep_H\Mpl^2H^2\,,\quad
    \Lambda=3\Mpl^2H^2\,,\quad
    \lambda=-I\ep_H\Mpl^2H\,,\quad
    M_2^4=-\fr{1}{6}I\ep_H\Mpl^2H^2\,,\\
    &m_1^2=m_2^2=\fr{1}{3}I\ep_H\Mpl^2\,,\quad
    m_3^4=I\ep_H\Mpl^2H^2\,,\quad
    m_4^3=\fr{2}{3}I\ep_H\Mpl^2H\,,
    \end{align}
at the leading order.
Here, $\ep_H\coloneqq -\dot{H}/H^2$ is the Hubble slow-roll parameter and $I\coloneqq 3\dot{\si}/(\ep_H H)$ characterizes the anisotropy.
This means that $H\la$, $M_2^4$, $H^2m_1^2$, $H^2m_2^2$, $m_3^4$, and $Hm_4^3$ are suppressed by the small anisotropy~$I$ compared to $c$ and $\Lambda$.

Having introduced the action in the unitary gauge, now we recover the NG
bosons ($\pi$ and $\eta_{\widehat m}$ for the broken time diffeomorphism
and the local rotation symmetry, respectively) via the St{\"u}ckelberg
trick. Note that $\chi$ is not the NG mode and we do not need
to recover the $U(1)$ symmetry, which is not spontaneously broken.
This can be achieved by the following replacements:
    \begin{align}
    t\to t+\pi\,,
    \quad
    e_\mu^{\bar3}\to e_\mu^{\widehat m}\eta_{\widehat m}+e_\mu^{\bar3}\brab{1+\fr{1}{2}\braa{\eta_0^2-\eta_1^2-\eta_2^2}}+\mO(\eta^3)\,.
    \end{align}
It should be noted that $\eta_0$ is nondynamical in the present case with $m_1^2=m_2^2$.
We study the case with dynamical $\eta_0$ in \S \ref{ssec:dynamicalh0}.
In the decoupling limit where the metric fluctuations are neglected, the quadratic action for the NG bosons takes the following form:
    \begin{align}
    S^{(2)}&=S_\pi+S_\eta+S_{\rm mix}\,, \\
    S_\pi&=\int d\tau d^3\ve{x}a^2\fr{f_\pi^4}{2}\brab{\pi'^2-c_\pi^2(\pa_A\pi)^2-\ti{c}_\pi^2(\pa_3\pi)^2}\,, \\
    S_\eta&=\int d\tau d^3\ve{x}a^2\biggl\{\fr{1}{2}\brab{-c_0^2(\pa_A\eta_0)^2-\ti{c}_0^2(\pa_3\eta_0)^2-m_0^2a^2\eta_0^2} \nonumber \\
    &~~~~~~~~~~~~~~~~~~~+\fr{f_\perp^2}{2}\brab{\eta_A'^2-c_{\perp}^2\fr{(\pa_A\eta_B-\pa_B\eta_A)^2}{2}-\ti{c}_\perp^2(\pa_3\eta_A)^2} \nonumber \\
    &~~~~~~~~~~~~~~~~~~~+g_0\eta_0\pa_A\eta^A{}'\biggr\}\,, \\
    S_{\rm mix}&=\int d\tau d^3\ve{x}a^2\pa_3\pi\braa{-g_1H\pa_A\eta^A-g_2\eta_0'-\ti{g}_2a\eta_0}\,,
    \end{align}
where the indices~$A$ and $B$ run over $1,2$ and a prime denotes a
derivative with respect to the conformal time~$\tau$. One easily
observes that there is no kinetic term of $\eta_0$ and that it is really
nondynamical. Note also that here we have used the slow-roll
approximation and taken the small-anisotropy limit. The coefficients are
given by
    \begin{align}
    &f_\pi^4=2(c+2M_2^4)\,,\quad
    c_\pi^2=\fr{2c+H^2m_1^2}{2(c+2M_2^4)}\,,\quad
    \ti{c}_\pi^2=\fr{2c-m_3^4}{2(c+2M_2^4)}\,, \\
    &c_0^2=\ti{c}_0^2=-m_1^2\,,\quad
    m_0^2=-m_3^4\,, \\
    &f_\perp^2=m_1^2\,,\quad
    c_{\perp}^2=\ti{c}_\perp^2=1\,, \\
    &g_0=m_1^2\,,\quad 
    g_1=m_1^2\,,\quad
    g_2=m_4^3\,,\quad
    \ti{g}_2=2Hm_4^3+m_3^4\,.
    \end{align}
Inserting our particular choices of the parameters given in
(\ref{eq:parametersfirst})--(\ref{eq:parameterslast}) yields
    \begin{align}
    S^{(2)}&=S_\pi+S_\eta+S_{\rm mix}\,, 
    \label{S2-eta0nonD}
    \\
    S_\pi&=\int d\tau \fr{d^3\ve{k}}{(2\pi)^3}\,\ep_H\Mpl^2a^2H^2\braa{\pi_\ve{k}'\pi_{-\ve{k}}'-k^2\pi_\ve{k}\pi_{-\ve{k}}}\,, 
    \\
    S_\eta&=\int d\tau \fr{d^3\ve{k}}{(2\pi)^3}\,\fr{1}{6}I\ep_H\Mpl^2a^2\biggl[k^2\eta_{0,\ve{k}}\eta_{0,-\ve{k}}+\eta_{1,\ve{k}}'\eta_{1,-\ve{k}}'-k_3^2\eta_{1,\ve{k}}\eta_{1,-\ve{k}}+\eta_{2,\ve{k}}'\eta_{2,-\ve{k}}'-k^2\eta_{2,\ve{k}}\eta_{2,-\ve{k}} 
    \nonumber \\
    & \hspace{12em} -2ik_1\eta_{0,\ve{k}}(\eta_{1,-\ve{k}}'+3aH\eta_{1,-\ve{k}})\biggr]\,, 
    \\
    S_{\rm mix}&=\int d\tau \fr{d^3\ve{k}}{(2\pi)^3}\,\fr{2}{3}I\ep_H\Mpl^2a^2Hk_3\pi_\ve{k}\braa{k_1\eta_{1,-\ve{k}}-i\eta_{0,-\ve{k}}'-5iaH\eta_{0,-\ve{k}}}\,. \label{lag1}
    \end{align}
Here, we have moved to the Fourier space and taken the wavenumber vector to be $\ve{k}=(k_1,0,k_3)$, which does not lose generality thanks to the symmetry in the $x$-$y$ plane.
Note also that we have omitted terms that are higher order in $\ep_H$ and $I$. 
We require $\ep_H>0$ and $I>0$ so that the kinetic terms of $\pi$ and $\eta_A$ have a correct sign.
We summarize some useful formulae in the slow-roll and small-anisotropy approximations in Appendix~\ref{AppB}.
The mass terms for $\pi$ and $\eta_A$ have been neglected because they are respectively proportional to $\ep_H^2$ and $I^2\ep_H^2$.
Since $\eta_0$ is nondynamical, we can eliminate it by use of its Euler-Lagrange equation.
Before doing so, we perform a gauge transformation~$\delta A_\mu\to \delta A_\mu+\pa_\mu\alpha$ such that $\delta A_3=0$ in the new gauge.
This can be achieved by choosing the Fourier component of $\alpha$ as
    \begin{equation}
    \alpha_\ve{k}=2i\fr{aH\bar{v}}{k_3}\pi_\ve{k}\,.
    \end{equation}
Accordingly, $\eta_0$ and $\eta_1$ are transformed respectively as
    \begin{equation}
    \begin{split}
    \eta_{0,\ve{k}} & \to \eta_{0,\ve{k}}+2i\fr{H}{k_3}(\pi_\ve{k}'+3aH\pi_\ve{k})\,, 
    \\
    \eta_{1,\ve{k}} & \to \eta_{1,\ve{k}}-2H\fr{k_1}{k_3}\pi_\ve{k}\,. 
    \end{split}
    \label{gauge}
    \end{equation}
Note that $\eta_2$ is unchanged due to our choice of the wavenumber vector.
Then, $S_{\rm mix}$ changes as
    \begin{align}
    S_{\rm mix}=\int d\tau \fr{d^3\ve{k}}{(2\pi)^3}\,\fr{1}{3}I\ep_H\Mpl^2a^2Hk_3\pi_\ve{k}\braa{k_1\eta_{1,-\ve{k}}-12iaH\eta_{0,-\ve{k}}}\,,
    \end{align}
while the other parts of the action remain unchanged\footnote{Precisely speaking, the transformation~\eqref{gauge} results in a nontrivial contribution on $S_\pi$, but it is of order $I\ep_H$ and thus can be neglected.}.
Now, the Euler-Lagrange equation for $\eta_0$ allows us to express $\eta_0$ itself as a function of $\pi$, $\eta_1$, and their derivatives.
In the Fourier space, we have
    \begin{equation}
    \eta_{0,\ve{k}}=-i\fr{k_1}{k^2}\braa{\eta_{1,\ve{k}}'+3aH\eta_{1,\ve{k}}}+12iaH^2\fr{k_3}{k^2}\pi_\ve{k}\,.
    \end{equation}
Substituting this back into the action, we are left with the following expression:
    \begin{align}
    S^{(2)}=\int d\tau \fr{d^3\ve{k}}{(2\pi)^3}
    \biggl[ \nonumber
    &\fr{1}{2}\hat{\pi}_\ve{k}'\hat{\pi}_{-\ve{k}}'-\fr{1}{2}\braa{k^2-2a^2H^2} \hat{\pi}_{\ve{k}}\hat{\pi}_{-\ve{k}} 
    \nonumber \\
    &+\fr{1}{2}\hat{\eta}_{1,\ve{k}}'\hat{\eta}_{1,-\ve{k}}'-\fr{1}{2}\braa{k^2-2a^2H^2} \hat{\eta}_{1,\ve{k}}\hat{\eta}_{1,-\ve{k}} 
    \nonumber \\
    &+\fr{1}{2}\hat{\eta}_{2,\ve{k}}'\hat{\eta}_{2,-\ve{k}}'-\fr{1}{2}\braa{k^2-2a^2H^2} \hat{\eta}_{2,\ve{k}}\hat{\eta}_{2,-\ve{k}} 
    \nonumber \\
    &+2\sqrt{6I}aH\sin\theta_\ve{k}\,\hat{\pi}_\ve{k}\hat{\eta}_{1,-\ve{k}}'
    +4\sqrt{6I}a^2H^2\sin\theta_\ve{k}\,\hat{\pi}_\ve{k}\hat{\eta}_{1,-\ve{k}}
    \biggr]\,,
    \end{align}
where we have defined the canonically normalized fields as
    \begin{align}
    \hat{\pi}_{\ve{k}} & \coloneqq \sqrt{2\ep_H}a\Mpl H\pi_{\ve{k}}\,,
    \\
    \hat{\eta}_{1,\ve{k}} & \coloneqq \sqrt{\fr{I\ep_H}{3}}a\Mpl\fr{|k_3|}{k}\eta_{1,\ve{k}}\,,
    \\
    \hat{\eta}_{2,\ve{k}} & \coloneqq \sqrt{\fr{I\ep_H}{3}}a\Mpl\eta_{2,\ve{k}}\,,
    \end{align}
and $\theta_\ve{k}$ denotes the angle measured from the $z$-axis to $\ve{k}$, i.e.
    \begin{equation}
    \sin \theta_\ve{k}=\fr{k_1k_3}{k|k_3|}\,,\quad
    \cos \theta_\ve{k}=\fr{k_3}{k}\,.
    \end{equation}
Thus, we see that the noninteracting part of the Lagrangian is that of free fields.
We then quantize $Q_0\coloneqq \hat{\pi}$, $Q_1\coloneqq \hat{\eta}_1$, and $Q_2\coloneqq \hat{\eta}_2$ as
    \begin{equation}
    Q_{n,\ve{k}}(\tau)=u_k(\tau) a_{n,\ve{k}}+u_k^\ast(\tau) a_{n,-\ve{k}}^\da\quad(n=0,1,2)\,,
    \end{equation}
with the standard commutation relation:
    \begin{equation}
    \Big[ a_{m,\ve{k}},a_{n,\ve{k}'}^\da \Big] = (2\pi)^3\delta_{mn}\delta^{(3)}(\ve{k}-\ve{k}')\,.
    \end{equation}
Here, the mode function~$u_k$ is given by
    \begin{equation}
    u_k(\tau)=\fr{1}{\sqrt{2k}}\braa{1-\fr{i}{k\tau}}e^{-ik\tau}\,, 
    \label{modef}
    \end{equation}
where we have used the de~Sitter approximation~$aH=(-\tau)^{-1}$.
The interaction Hamiltonian is given by
    \begin{equation}
    H_I=-2\sqrt{6I}\int \fr{d^3\ve{k}}{(2\pi)^3}\,\sin\theta_\ve{k}
    \brab{(-\tau)^{-1}\hat{\pi}_\ve{k}\hat{\eta}_{1,-\ve{k}}'
    +2(-\tau)^{-2}\hat{\pi}_\ve{k}\hat{\eta}_{1,-\ve{k}}}\,. 
    \label{Hint1}
    \end{equation}
By use of the standard in-in formalism computation
to compute the expectation value of an operator $O(\tau)$,
    \begin{align}
    \Braket{{\rm in}|O(\tau)|{\rm in}}=&\Braket{0|O(\tau)|0}
    +i\int_{\tau_i}^\tau d\tau_1 \Braket{0|[H_I(\tau_1),O(\tau)]|0} 
    \nonumber \\
    &+i^2\int_{\tau_i}^\tau d\tau_1 \int_{\tau_i}^{\tau_1} d\tau_2 \Braket{0|[H_I(\tau_2),[H_I(\tau_1),O(\tau)]]|0}+\cdots\,, \label{in-in}
    \end{align}
the $\hat{\pi}$-$\hat{\pi}$ correlation function can be calculated as
    \begin{equation}
    \Braket{{\rm in}|\hat{\pi}_\ve{k}(\tau)\hat{\pi}_{\ve{k}'}(\tau)|{\rm in}}
    =(2\pi)^3\delta^{(3)}(\ve{k}+\ve{k}')|u_k(\tau)|^2\brab{1+\Delta_{\rm aniso}(\ve{k})}\,.
    \end{equation}
Here, $\Delta_{\rm aniso}(\ve{k})$ is the anisotropic part given by
    \begin{equation}
    \Delta_{\rm aniso}(\ve{k})=\fr{192I\sin^2\theta_\ve{k}}{|u_k(\tau)|^2}
    \int_{\tau_i}^\tau d\tau_1 \int_{\tau_i}^{\tau_1} d\tau_2
    \Im \Big[ u_k(\tau_1)u_k^\ast(\tau) \Big]
    \Im \Big[ u_k(\tau_2)u_k^\ast(\tau)v_k(\tau_2)v_k^\ast(\tau_1) \Big] \,, 
    \label{Daniso}
    \end{equation}
with $v_k(\tau)\coloneqq (-\tau)^{-1}u_k'(\tau)+2(-\tau)^{-2}u_k(\tau)$ and $\Im$ denoting the imaginary part of a complex number.
Assuming that the mode is superhorizon (i.e. $|k\tau|\ll 1$), the main contribution to the integral in \eqref{Daniso} comes from the region where $|k\tau_2|\ll 1$.
In this limit, $|u_k(\tau)|^2 \approx 1/(2k^3\tau^2)$ and the integrand in~\eqref{Daniso} can be approximated as
    \begin{align}
    \Im \Big[ u_k(\tau_1)u_k^\ast(\tau) \Big]
    \Im \Big[ {u_k(\tau_2)u_k^\ast(\tau)v_k(\tau_2)v_k^\ast(\tau_1)} \Big]
    \approx 
    \fr{1}{8k^3\tau^2}\fr{(\tau^3-\tau_1^3)(\tau^3-\tau_2^3)}{\tau_1^4\tau_2^4}\,.
    \end{align}
Thus, we have
    \begin{align}
    \Delta_{\rm aniso}(\ve{k}) 
    & \approx 
    48I\sin^2\theta_\ve{k}\int_{-1/k}^\tau d\tau_1 \int_{-1/k}^{\tau_1} d\tau_2\fr{(\tau^3-\tau_1^3)(\tau^3-\tau_2^3)}{\tau_1^4\tau_2^4}
    \nonumber\\
    & \approx
    24I\sin^2\theta_\ve{k} N_k^2
    \eqqcolon g_* \sin^2\theta_\ve{k}\,, 
    \label{Daniso1}
    \end{align}
with $N_k\coloneqq -\ln(-k\tau)$ being the $e$-folding number from the horizon exit to the end of inflation.
Since $I>0$, \eqref{Daniso1} predicts an oblate anisotropy.
This is consistent with the result of~\cite{Dulaney:2010sq,Gumrukcuoglu:2010yc,Watanabe:2010fh}.
For a large $N_k$, we need a more careful analysis~\cite{Naruko:2014bxa}.
Moreover, we can extend the analysis to the bispectrum~\cite{Bartolo:2012sd}.

Let us comment on observational constraints on the anisotropy.
Regarding the background anisotropy, it is known to satisfy $\dot{\sigma}/H_0<10^{-9}$~\cite{Bunn:1996ut}. 
On the other hand, the equations~$\dot{\sigma}/H=I\epsilon_H/3$ and $g_*=24IN_k^2$ tell us $\dot{\sigma}/H\lesssim 5\times 10^{-8}\epsilon_H$ in the inflationary era for $N_k=50$ and $g_*<0.01$.
For instance, for $\epsilon_H=0.01$, we expect $\dot{\sigma}/H$ is at most of order~$10^{-10}$.
Given that the energy density of the vector field decays after the inflation, the background anisotropy decreases as $\dot{\sigma}/H\propto a^{-1}$, and thus the aforementioned bound on $\dot{\sigma}/H_0$ can be easily satisfied.
Even if the vector field does not decay, the background anisotropy decreases with time during the reheating epoch, so that it does not violate the observational constraint.

\subsection{Dynamical $\eta_0$}
\label{ssec:dynamicalh0}

Now we proceed to the case where $\eta_0$ is dynamical.
In order to illustrate the key ideas while avoiding technical complications, we consider the following action:
    \begin{align}
    S=\int d^4x\sqrt{-g}\biggl[&
    \fr{\Mpl^2}{2}R+(c-\Lambda)-c\delta g^{00}+\la D_\mn e^{\mu{\bar3}}n^\nu
    -\fr{m_1^2}{2}D_\mn D^\mn
    +\fr{m_3^4}{2}\braa{n^\mu e_\mu^{\bar3}}^2
    -\fr{m_9^2}{2}\braa{D_\mn n^\nu}^2
    \biggr]\,, \label{SeffUGm9}
    \end{align}
where $c$, $\Lambda$, $\la$, $m_1$, $m_3$, and $m_9$ are assumed to be constant.
In this case, the quadratic action for the NG bosons takes the following form:
    \begin{align}
    S^{(2)}&=S_\pi+S_\eta+S_{\rm mix}\,, 
    \\
    S_\pi&=\int d\tau \fr{d^3\ve{k}}{(2\pi)^3}\,a^2\fr{f_\pi^4}{2}\braa{\pi_\ve{k}'\pi_{-\ve{k}}'-k^2\pi_\ve{k}\pi_{-\ve{k}}}\,, 
    \\
    S_\eta&=\int d\tau \fr{d^3\ve{k}}{(2\pi)^3}\,a^2\biggl\{
    \fr{f_0^2}{2}\braa{\eta_{0,\ve{k}}'\eta_{0,-\ve{k}}'-k^2\eta_{0,\ve{k}}\eta_{0,-\ve{k}}} 
    \nonumber \\
    & \hspace{8em}
    +\fr{f_\perp^2}{2}\brab{\eta_{1,\ve{k}}'\eta_{1,-\ve{k}}'-k^2\eta_{1,\ve{k}}\eta_{1,-\ve{k}}
    +\eta_{2,\ve{k}}'\eta_{2,-\ve{k}}'-(2k_1^2+k_3^2)\eta_{2,\ve{k}}\eta_{2,-\ve{k}}} 
    \nonumber \\
    & \hspace{8em}
    -ig_0aHk_1\eta_{0,\ve{k}}\eta_{1,-\ve{k}}
    \biggr\}\,, 
    \\
    S_{\rm mix}&=\int d\tau \fr{d^3\ve{k}}{(2\pi)^3}\,a^2k_3\pi_\ve{k}\braa{g_1Hk_1\eta_{1,-\ve{k}}-ig_2a\eta_{0,-\ve{k}}}\,,
    \end{align}
where we have moved to the Fourier space and chosen $\ve{k}=(k_1,0,k_3)$ as in the previous section.
Here, we have taken $m_3^4=H^2(m_1^2+2m_9^2)-H\la$ so that the mass term for $\eta_0$ vanishes.
Note also that we have assumed that the terms~$H\la$, $H^2m_1^2$, and $H^2m_9^2$ are negligibly small compared to $c$.
The coefficients are given by
    \begin{equation}
    \begin{split}
    &f_\pi^4=2c\,,\quad
    f_0^2=m_9^2-m_1^2\,,\quad
    f_\perp^2=m_1^2\,, 
    \\
    &g_0=m_1^2-\fr{\la}{H}\,,\quad
    g_1=m_1^2+\fr{\la}{H}\,,\quad
    g_2=-2H\la+2H^2m_9^2\,.
    \end{split} \label{Seff_param}
    \end{equation}
Thus, one can circumvent ghost and gradient instability by choosing the parameters such that
    \begin{align}
    c>0
    \quad \text{and} \quad
    m_9^2>m_1^2>0\,.
    \end{align}
Thus, the presence of $m_9^2$ term circumvents the no-go theorem~\cite{Himmetoglu:2008zp,Himmetoglu:2008hx}.
In the Fourier space, the quadratic action can be written as
    \begin{align}
    S^{(2)}&=\int d\tau \fr{d^3\ve{k}}{(2\pi)^3}\,\biggl[
    \fr{1}{2}\hat{\pi}_\ve{k}'\hat{\pi}_{-\ve{k}}'-\fr{1}{2}(k^2-2a^2H^2)\hat{\pi}_\ve{k}\hat{\pi}_{-\ve{k}}
    +\fr{1}{2}\hat{\eta}_{0,\ve{k}}'\hat{\eta}_{0,-\ve{k}}'-\fr{1}{2}(k^2-2a^2H^2)\hat{\eta}_{0,\ve{k}}\hat{\eta}_{0,-\ve{k}} 
    \nonumber \\
    & \hspace{7em}
    +\fr{1}{2}\hat{\eta}_{1,\ve{k}}'\hat{\eta}_{1,-\ve{k}}'-\fr{1}{2}(k^2-2a^2H^2)\hat{\eta}_{1,\ve{k}}\hat{\eta}_{1,-\ve{k}}
    +\fr{1}{2}\hat{\eta}_{2,\ve{k}}'\hat{\eta}_{2,-\ve{k}}'-\fr{1}{2}(2k_1^2+k_3^2-2a^2H^2)\hat{\eta}_{2,\ve{k}}\hat{\eta}_{2,-\ve{k}} \nonumber \\
    & \hspace{7em}
    -i\fr{g_0}{f_0f_\perp}aHk_1\hat{\eta}_{0,\ve{k}}\hat{\eta}_{1,-\ve{k}}
    +\fr{g_1}{f_\pi^2f_\perp}Hk_1k_3\hat{\pi}_\ve{k}\hat{\eta}_{1,-\ve{k}}
    -i\fr{g_2}{f_\pi^2f_0}ak_3\hat{\pi}_\ve{k}\hat{\eta}_{0,-\ve{k}}
    \biggr]\,,
    \end{align}
where the fields have been canonically normalized by
    \begin{align}
    \hat{\pi}_{\ve{k}} & \coloneqq af_\pi^2 \pi_{\ve{k}}\,,
    \\
    \hat{\eta}_{0,\ve{k}} & \coloneqq af_0 \eta_{0,\ve{k}}\,,
    \\
    \hat{\eta}_{A,\ve{k}} & \coloneqq af_\perp \eta_{A,\ve{k}}\quad (A=1,2)\,.
    \end{align}

Let us now calculate the anisotropic part of the $\hat{\pi}$-$\hat{\pi}$ correlation function.
For simplicity, we assume $m_1^2/m_9^2\ll 1$ and $\la/(Hm_9^2)\ll 1$.
Then, the main contribution comes from the interaction between $\hat{\pi}$ and $\hat{\eta}_0$, and thus we hereafter omit $\hat{\eta}_1$ and $\hat{\eta}_2$.
As in the previous case of nondynamical $\eta_0$, the noninteracting part of the Lagrangian is that of free fields, so that the mode functions for $\hat{\pi}$ and $\hat{\eta}_0$ are again given by~\eqref{modef}.
The interaction Hamiltonian in the present case is
    \begin{equation}
    H_I=i\fr{g_2}{f_\pi^2f_0H}(-\tau)^{-1}\int \fr{d^3\ve{k}}{(2\pi)^3}\,k\cos\theta_\ve{k}\hat{\pi}_\ve{k}\,\hat{\eta}_{0,-\ve{k}}\,, 
    \label{Hint2}
    \end{equation}
where we have used $k_3=k\cos\theta_\ve{k}$.
With the aid of \eqref{in-in}, the anisotropic part of the $\hat{\pi}$-$\hat{\pi}$ correlation function is given by
    \be
    \Delta_{\rm aniso}(\ve{k})=\fr{1}{2}\braa{\fr{g_2}{f_\pi^2f_0H}}^2\cos^2\theta_\ve{k}
    \brac{\left|\int_{-k\tau}^\infty \fr{dx_1}{x_1}\ti{u}(x_1)^2\right|^2
    -2\,\Re\brab{\fr{\ti{u}(-k\tau)^2}{|\ti{u}(-k\tau)|^2}\int_{-k\tau}^\infty\fr{dx_1}{x_1}|\ti{u}(x_1)|^2\int_{x_1}^\infty\fr{dx_2}{x_2}\ti{u}^\ast(x_2)^2}}\,,
    \ee
where $\Re$ denotes the real part of a complex number and we have defined
    \be
    \ti{u}(x)\coloneqq\braa{1+\fr{i}{x}}e^{ix}\,.
    \ee
Evaluating the integral at the leading order, we have
    \be
    \Delta_{\rm aniso}(\ve{k})=\braa{\fr{\pi^2}{6}-\fr{1}{4}}\braa{\fr{g_2}{f_\pi^2f_0H}}^2\cos^2\theta_\ve{k}+\mO(-k\tau)\,. \label{Daniso2}
    \ee
Note that, in contrast to~\eqref{Daniso1}, there is no enhancement factor~$N_k^2$.
This difference should come from the form of the interaction Hamiltonian, \eqref{Hint1} and \eqref{Hint2}, where we have a derivative interaction in \eqref{Hint1} but not in \eqref{Hint2}.
Note also that \eqref{Daniso2} predicts a prolate anisotropy.


\section{Conclusions}
\label{sec:conclusions}

By extending the idea of the EFT approach in isotropic inflation introduced in~\cite{Cheung:2007st}, we have developed an EFT of anisotropic inflation. 
The point is that we have introduced three NG bosons~$\eta_{\widehat m}$ for the broken spatial rotation symmetry in addition to $\pi$ for the broken time translation symmetry.
After studying the simplest model~\eqref{simplest}, we have constructed the effective action in the unitary gauge up to quadratic order in the perturbations. 
The building blocks of the effective action in the unitary gauge are the metric~$g_\mn$, the timelike unit vector~$n_\mu$, the $\bar{3}$-component of the vierbein~$e^{\bar{3}}_\mu$, and their derivatives.
The lowest few terms in derivatives are presented in~\eqref{SeffUG}.
Based on the so-obtained effective action, we have computed the statistical anisotropy in the scalar two-point correlation function.
For the conventional model~\eqref{SeffUGsimple} having nondynamical $\eta_0$, the anisotropy is given by~\eqref{Daniso1}, which reproduces the well-known result of~\cite{Watanabe:2010fh}.
On the other hand, for the model~\eqref{SeffUGm9} with dynamical $\eta_0$, the anisotropy is given by~\eqref{Daniso2}, which is in sharp contrast to the conventional case in the sense that the enhancement factor~$N_k^2$ is absent.

It is useful to compare our approach with those in the literature~\cite{Abolhasani:2015cve,Rostami:2017wiy}.  The authors of \cite{Rostami:2017wiy} studied interactions between the NG boson for the broken time diffeomorphism, i.e., the inflaton fluctuations, and a $U(1)$ gauge field which sources the anisotropy. The same class of models was reformulated in \cite{Abolhasani:2015cve} in terms of four NG bosons associated with time and spatial diffeomorphisms. In particular, the gauge field fluctuations were described by three NG bosons for spatial diffeomorphisms~$\pi^i$. On the other hand, our approach identifies the gauge field fluctuations with NG bosons for broken local rotations~$\eta_{\widehat{m}}$. These two languages can be translated into each other by an identification~$\eta\sim\partial\pi$. Note that our NG boson~$\eta_{\widehat{m}}$ is directly related to the gauge field fluctuations without derivatives (see also discussion in \S \ref{sec:example}). Besides, our approach is not restricted to the models with gauge symmetry. For example, as we illustrated in \S \ref{ssec:dynamicalh0}, our approach can capture models with dynamical $\eta_0$.

It is also interesting to mention the relation between the present case and the case of broken boost invariance studied in \cite{ArmendarizPicon:2010mz,Delacretaz:2015edn}.
In the latter case, the construction of the effective action is similar to the present case of broken rotation invariance, but it is not a mere analytical continuation:~For example, as \eqref{SeffUGm9} and \eqref{Seff_param} in the present paper show, the choice $m_1^2>0$ gives a wrong sign of the kinetic term of $\eta_0$, while $\eta_{1,2}$ have a correct sign. 
Therefore, we need to arrange other operators (such as $m_9^2$) carefully to avoid ghost instabilities. 
On the other hand, in the case of boost breaking, the operator analogous to the present $m_1^2$ gives the same sign of the kinetic term of three NG bosons~$\eta_{1,2,3}$ for broken boosts, so that we do not need to worry about such potential ghost instabilities. 
Besides, the boost breaking case has the same spacetime symmetry as the standard isotropic inflation, so it does not lead to any anisotropy (both in the background and fluctuations). 
Hence, phenomenological consequences are completely different from the present case.

In the present paper, we computed only the scalar two-point correlation function. 
It would be also intriguing to calculate the correlation function for the tensor modes and the cross-correlation between the scalar and tensor modes.
The cross-correlation gives rise to the $TB$ correlation in the CMB~\cite{Watanabe:2010bu}.
Moreover, for the simplest model~\eqref{simplest}, there exist consistency relations between these correlation functions~\cite{Soda:2012zm}.
If one could find such consistency relations for a broader class of models described by our EFT framework, they allow us to give a model-independent test of anisotropic inflation.
We leave these issues for future study.

It should be stressed that our formalism allows us to go beyond anisotropic inflation. Indeed, we can discuss inflation with spontaneous breaking of the translation invariance, boost invariance, and a combination of these
symmetry breaking patterns.


\acknowledgments{J.~G.\ is supported by the Basic Science Research
Program (2016R1D1A1B03930408) and Mid-career Research Program
(2019R1A2C2085023) through the National Research Foundation of Korea
Research Grants.  J.~G.\ also acknowledges the Korea-Japan Basic Scientific
Cooperation Program supported by the National Research Foundation of
Korea and the Japan Society for the Promotion of Science
(2018K2A9A2A08000127).  T.~N.\ is supported in part by JSPS KAKENHI Grant
Numbers JP17H02894 and JP18K13539, and MEXT KAKENHI Grant Number
JP18H04352.  G.~S.\ is supported in part by the DOE grant DE-SC0017647
and the Kellett Award of the University of Wisconsin.  J.~S.\ is in part
supported by JSPS KAKENHI Grant Numbers JP17H02894, JP17K18778,
JP15H05895, JP17H06359, JP18H04589.  J.~S.\ and T.~N.\ are also supported
by JSPS Bilateral Joint Research Projects (JSPS-NRF collaboration) ``String Axion Cosmology.''  K.~T.\ is supported in part by JSPS KAKENHI
Grant Numbers~JP17H02894 and JP17K18778. M.~Y.\ is supported in part
by JSPS Grant-in-Aid for Scientific Research Numbers 18K18764, MEXT KAKENHI
Grant-in-Aid for Scientific Research on Innovative Areas Numbers 15H05888,
18H04579, Mitsubishi Foundation, JSPS and NRF under the Japan-Korea
Basic Scientific Cooperation Program, and JSPS Bilateral Open
Partnership Joint Research Projects.}


\appendix

\section{A brief review of~\cite{Hidaka:2014fra}}
\label{App:review}

In this Appendix, we briefly review \cite{Hidaka:2014fra} to demonstrate how the viewpoint of local symmetries is useful to identify the NG field associated with spontaneous spacetime symmetry breaking.

\subsection{Local decomposition of spacetime symmetries}

We begin by summarizing basics of spacetime symmetries and transformation rules of local fields. Consider a spacetime symmetry associated with a coordinate transformation,
\begin{align}
\label{general_coord}
x^\mu\to x'^\mu=x^\mu-\epsilon^\mu(x)\,.
\end{align}
Here and in what follows, we focus on infinitesimal transformations, but generalization to finite transformations is straightforward. To determine the transformation rule of a local field located at $x^\mu=x_*^\mu$, it is convenient to expand the transformation parameter~$\epsilon^\mu(x)$ covariantly as
\begin{align}
\epsilon^\mu(x)=\epsilon^\mu(x_*)+(x^\nu-x_*^\nu)\nabla_\nu\epsilon^\mu(x_*)+\mathcal{O}\Big((x-x_*)^2\Big)\,,
\end{align}
where the first term is the zeroth order in $x-x_*$
and
describes translation of the coordinate system.
On the other hand,
deformations of the coordinate frame are encoded in the second term
(the linear order in $x-x_*$),
which can be decomposed~as
\begin{align}
\label{general_decomposition}
\nabla_\mu\epsilon^\nu
=\delta_\mu^\nu\,\lambda+s_\mu{}^\nu+\omega_\mu{}^\nu\,,
\quad
{\rm with}
\quad
s_{\mu}{}^\mu=0\,,
\quad
s_{\mu\nu}=s_{\nu\mu}\,,
\quad
\omega_{\mu\nu}=-\omega_{\nu\mu}\,.
\end{align}
Here, the trace part~$\lambda$ and the symmetric traceless part~$s_{\mu\nu}$
are local isotropic rescalings (dilatations)
and local anisotropic rescalings,
respectively.
The antisymmetric part $\omega_{\mu\nu}$ corresponds to
local Lorentz transformations. In particular, the metric field transforms as
\begin{align}
\delta g_{\mu\nu}=\nabla_\mu\epsilon_\nu+\nabla_\nu\epsilon_\mu
=2g_{\mu\nu}\lambda+2s_{\mu\nu}\,,
\end{align}
so that isometric transformations correspond to $\lambda =s_{\mu\nu}=0$. For simplicity, we focus on isometric transformations in the following (see~\cite{Hidaka:2014fra} for more general cases).

Then, the transformation rule of a local field is determined by its spin.
When a field~$\Phi(x)$ is in a representation~$\Sigma_{mn}$
of the Lorentz algebra, its transformation rule is given in terms of $\epsilon^\mu(x)$ and $\omega_{\mu\nu}(x)$ as
\begin{align}
\label{decomposition1}
\delta \Phi=\frac{1}{2}\omega^{mn}(x)\Sigma_{mn}\Phi(x)+\epsilon^\mu(x)\nabla_\mu\Phi(x)\,.
\end{align}
Here, the curved-spacetime indices
(Greek letters)
and the local-Lorentz indices
(Latin letters)
are converted
by the vierbein $e_\mu^m$ as
$\omega^{mn}=e^m_\mu e_\nu^n\omega^{\mu\nu}$.
The covariant derivative $\nabla_\mu\Phi$ is defined
in terms of the spin connection $S_\mu^{mn}$ as
\begin{align}
\label{covariant_derivative}
\nabla_\mu\Phi=\partial_\mu\Phi+\frac{1}{2}S_\mu^{mn}\Sigma_{mn}\Phi\,,
\quad
\text{with}
\quad
S_\mu^{mn}=e_\nu^m\partial_\mu e^{\nu n}
+e_\lambda^m\,\Gamma^\lambda_{\mu\nu}\,e^{\nu n}\,.
\end{align}
As we explain in the next subsection, it is convenient to decompose the transformation~\eqref{decomposition1} into diffeomorphisms and local Lorentz transformations, when identifying broken symmetries and the corresponding NG fields. For this purpose, we rewrite it as
\begin{align}
\label{decomposition2}
\delta\Phi&=\frac{1}{2}\left(\omega^{mn}+\epsilon^\mu S_{\mu}^{mn}\right)\Sigma_{mn}\Phi+\epsilon^\mu\partial_\mu\Phi\,.
\end{align}
Since the transformation rule of $\Phi$, $g_{\mu\nu}$, and $e_\mu^m$ under diffeomorphism and local Lorentz symmetries is given by
\begin{align}
\label{diffs}
\text{diffeomorphism}:&\quad\delta \Phi=\tilde{\epsilon}^\mu\partial_\mu\Phi\,,
\quad
\delta g_{\mu\nu}=\nabla_\mu\tilde{\epsilon}_\nu+\nabla_\nu\tilde{\epsilon}_\mu\,,
\quad
\delta e^m_\mu=\nabla_\mu\tilde{\epsilon}^m-\tilde{\epsilon}^\nu S_{\nu }^m{}_n\, e^n_\mu\,,
\\
\label{local_Lorentz}
\text{local Lorentz}:&\quad\delta \Phi=\frac{1}{2}\tilde{\omega}^{mn}\Sigma_{mn}\Phi\,,
\quad
\delta g_{\mu\nu}=0\,,
\quad
\delta e^m_\mu=\tilde{\omega}^m{}_n e^n_\mu\,,
\end{align}
we can embed the transformation \eqref{decomposition2} into these local symmetry transformations by the parameter choice
\begin{align}
\label{embed_relativistic}
\tilde{\epsilon}^\mu=\epsilon^\mu\,,
\quad
\tilde{\omega}^{mn}=\omega^{mn}+\epsilon^\mu S_\mu^{mn}\,.
\end{align}
Note that
the metric $g_{\mu\nu}$ and the vierbein $e_\mu^m$
are invariant under the isometric transformation, i.e., the parameter set~\eqref{embed_relativistic} with $\nabla_\mu\epsilon_\nu=-\nabla_\nu\epsilon_\mu=\omega_{\mu\nu}$, even though they are not invariant under general local symmetry transformations.

\subsection{Identification of broken symmetries and NG fields}

Next we discuss identification of broken symmetries and the corresponding NG fields. We say that a symmetry (of the action) is spontaneously broken when a condensation (vacuum expectation value) of a field in the linear representation~\eqref{decomposition2} is not invariant under the symmetry transformation:
\begin{align}
\langle\delta\Phi\rangle\neq0\,.
\end{align}
More concretely, the left-hand side is given by
\begin{align}
\label{vev_global}
\langle\delta\Phi\rangle
=\frac{1}{2}\left(\omega^{mn}+\epsilon^\mu S_{\mu}^{mn}\right)\Sigma_{mn}\langle\Phi\rangle+\epsilon^\mu\partial_\mu\langle\Phi\rangle
\,,
\end{align}
where the first term can be nonzero when the condensation has a nonzero spin. On the other hand, the second term can be nonzero when the condensation is inhomogeneous.

As an illustrative example, let us consider a non-gravitational system on the Minkowski spacetime,
\begin{align}
g_{\mu\nu}=\eta_{\mu\nu}\,,
\quad
e_\mu^m=\delta_\mu^m\,,
\end{align}
and two different types of condensations:
\begin{enumerate}
\item Time-dependent scalar condensation

Suppose that a scalar field $\phi$ has a time-dependent condensation:
\begin{align}
\langle\phi\rangle=\bar{\phi}(t)\,.
\end{align}
Its isometric transformation is then given by
\begin{align}
\langle\delta\phi\rangle=\epsilon^\mu\partial_\mu\langle\phi\rangle=\epsilon^0\dot{\bar{\phi}}\,,
\end{align}
so that the time-translational symmetry ($\epsilon^0=\text{const}$) and the boost symmetry ($\epsilon^0=b^ix^i$, $\epsilon^i=b^i t$ [$i=1,2,3$] with infinitesimal constant parameters $b^i$) are spontaneously broken. One would identify the scalar fluctuation as the NG field for the broken time-translational symmetry. However, there exists no dynamical degrees of freedom associated with the broken boosts. This illustrates the well-known mismatch between the number of broken (global) symmetries and that of NG fields for spacetime symmetries.

\item Time-dependent vector condensation

Actually, the same symmetry breaking pattern appears when a vector field $V^m$ has a time-dependent condensation,
\begin{align}
\langle V^m\rangle=\delta^m_0\bar{V}(t)\,,
\end{align}
along the time-direction. Its isometric transformation is given by
\begin{align}
\langle\delta V^m\rangle=\frac{1}{2}\omega^m{}_n\langle V^n\rangle+\epsilon^\mu\partial_\mu\langle V^m\rangle
=\frac{1}{2}\omega^m{}_0\bar{V}+\epsilon^0\delta^m_0\dot{\bar{V}}\,,
\end{align}
so that the time-translational and boost symmetries are spontaneously broken, just as the previous scalar example. One would identify the fluctuation of $V^0$ with the NG field for broken time-translation and $V^i$ as those for broken boosts. In contrast to the scalar example, there exist dynamical degrees of freedom associated with broken boosts\footnote{
Note that the modes associated with the broken boosts are generically gapped as is known as the inverse Higgs effects~\cite{Ogievetsky}. However, these gapped modes may affect low-energy dynamics if the gap is small enough compared to the energy scale of one's interests. Therefore, one needs to distinguish the scalar and vector condensations to capture the dynamics in this regime.}.

\end{enumerate}
So far, we have demonstrated that there is no one-to-one correspondence between broken spacetime symmetries and dynamical degrees of freedom in general, at least as long as we employ the viewpoint of global symmetries. Below we show that the viewpoint of local symmetries resolves the mismatch.

For this purpose, let us turn on the metric and vierbein to gauge the isometries into diffeomorphism and local Lorentz symmetries. Under each gauge transformation, the condensation transforms as
\begin{align}
\text{diffeomorphism}:&\quad\langle\delta \Phi\rangle=\tilde{\epsilon}^\mu\partial_\mu\langle\Phi\rangle\,,
\\
\text{local Lorentz}:&\quad\langle\delta \Phi\rangle=\frac{1}{2}\tilde{\omega}^{mn}\Sigma_{mn}\langle\Phi\rangle\,,
\end{align}
where the first and second terms in \eqref{vev_global} are nicely separated as a consequence of gauging. Now we interpret that the diffeomorphism symmetries are spontaneously broken when the condensation is inhomogeneous, whereas the local Lorentz symmetries are spontaneously broken when the condensation has a nonzero spin. In this picture, the previous two examples can be distinguished clearly. The identification of the corresponding NG fields is also straightforward. In particular, the NG fields for broken diffeomorphism symmetries can be eaten by the metric, whereas those for broken local Lorentz symmetries can be eaten by the vierbein to make it dynamical.

\section{Background equations of motion}
\label{AppA}

\subsection{Background geometry}

With the background metric \eqref{eq:BGmetric} and the background vierbein \eqref{background_real},
the nonzero components of the Christoffel symbols are
\begin{align}
\Gamma^0_{11}=(H+\dot{\sigma})a^2e^{2\sigma}\,,
\quad
\Gamma^0_{33}=(H-2\dot{\sigma})a^2e^{-4\sigma}\,,
\quad
\Gamma^1_{01}
=H+\dot{\sigma}\,,
\quad
\Gamma^3_{03}
=H-2\dot{\sigma}\,.
\end{align}
The spin connection defined by
\begin{align}
S_\mu^{mn}=e_\nu^m\partial_\mu e^{\nu n}+e^m_\lambda\Gamma^\lambda_{\mu\nu}e^{\nu n}\,,
\end{align}
has the following nonzero components:
\begin{align}
S_1^{\bar0\bar1}=-S_1^{\bar1\bar0}=(H+\dot{\sigma})ae^{\sigma}\,.
\quad
S_3^{\bar0\bar3}=-S_3^{\bar3\bar0}=(H-2\dot{\sigma})ae^{-2\sigma}\,.
\end{align}
The Ricci tensor,
\begin{align}
R_{\mu\nu}=\partial_\rho\Gamma^\rho_{\mu\nu}-\partial_\nu\Gamma^\rho_{\rho\mu}+\Gamma^\rho_{\rho\lambda}\Gamma^\lambda_{\mu\nu}-\Gamma^\rho_{\lambda\mu}\Gamma^\lambda_{\rho\nu}\,,
\end{align}
has the nonvanishing components:
\begin{align}
R_{00}&=-3H^2-3\dot{H}-6\dot{\sigma}^2\,,
\\
R_{11}&=a^2e^{2\sigma}\left(
3H^2+\dot{H}+3H\dot\sigma+\ddot{\sigma}
\right)\,,
\\
R_{33}&=
a^2e^{-4\sigma}\left(
3H^2+\dot{H}-6H\dot\sigma-2\ddot{\sigma}
\right)\,,
\end{align}
from which we obtain the Ricci scalar as
\begin{align}
R=12H^2+6\dot{H}+6\dot{\sigma}^2\,.
\end{align}
Note that the contribution of $\sigma$ to the Ricci scalar is always positive.
Finally, the nonzero components of the Einstein tensor, $\displaystyle G_{\mu\nu}=R_{\mu\nu}-\frac{1}{2}g_{\mu\nu}R$, are given by
\begin{align}
G_{00}&=3H^2-3\dot\sigma^2\,,
\\
G_{11}&=-a^2e^{2\sigma}\left(
3H^2+2\dot{H}-3H\dot{\sigma}-\ddot{\sigma}+3\dot{\sigma}^2
\right)\,,
\\
G_{33}&=-a^2e^{-4\sigma}\left(
3H^2+2\dot{H}+6H\dot{\sigma}+2\ddot{\sigma}+3\dot{\sigma}^2
\right)\,.
\end{align}

\subsection{Useful formulae}

To solve the background equations of motion,
it is convenient to introduce quantities
which are covariant under the unbroken symmetries
and vanish on the background.
Here, we introduce such quantities
for covariant derivatives of the ${\bar3}$-component
of vierbein.
Using the relation,
\begin{align}
\nabla_\mu e_\nu^{\bar3}=e_{\nu m}\nabla_\mu\delta^m_{\bar3}
=e_{\nu m}S_{\mu}^{m\bar{3}}\,,
\end{align}
we can compute its background value as
\begin{align}
\nabla_\mu e_\nu^{\bar3}=
-\delta_\mu^3\delta_\nu^0(H-2\dot{\sigma})ae^{-2\sigma}
=e_\mu^{\bar3}n_\nu(H-2\dot{\sigma})\,,
\end{align}
where $n_\mu=-\delta_\mu^0/\sqrt{-g^{00}}$
is the timelike unit vector.
Then,
if we introduce $D_{\mu\nu}$ as \eqref{eq:Dmn}
it is covariant under the unbroken symmetries
and vanish on the background.
For later use,
let us note several relations below.

First, the trace of $D_{\mu\nu}$ is given by
\begin{align}
D^\mu{}_\mu=\nabla^\mu e_\mu^{\bar3}-(H-2\dot{\sigma})n^\mu e_\mu^{\bar3}\,.
\end{align}
Since $e_0^{\bar{3}}$ vanishes on the background,
it suggests that
\begin{align}
\label{n^me_m}
\nabla^\mu e_\mu^{\bar3}=\mathcal{O}(\delta)\,.
\end{align}
Second,
if we contract $D_{\mu\nu}$ with $n^\mu e^{\nu\bar3}$,
we have
\begin{align}
\label{D_mnn^me^n}
D_{\mu\nu}n^\mu e^{\nu\bar3}=
\left[\nabla_\mu e_\nu^{\bar3}-e_\mu^{\bar3}n_\nu(H-2\dot{\sigma})\right]
n^\mu e^{\nu\bar3}
=-(H-2\dot{\sigma})\left(n^\mu e_\mu^{\bar3}\right)^2
=\mathcal{O}(\delta^2)\,.
\end{align}
The contraction with $e^{\mu\bar3}n^\nu$
is also given by
\begin{align}
\nonumber
D_{\mu\nu} e^{\mu\bar3}n^\nu
&=\left(\nabla_\mu e_\nu^{\bar3}\right) e^{\mu\bar3}n^\nu
+(H-2\dot{\sigma})
\\
\nonumber
&=\nabla_\mu\left( n^\nu e_\nu^{\bar3} e^{\mu\bar3}\right)
-n^\nu e_\nu^{\bar3}\left(\nabla^\mu e_\mu^{\bar3}\right)
-e^{\mu\bar3}e^{\nu\bar3}\nabla_\mu n_\nu
+(H-2\dot{\sigma})
\\
&=\nabla_\mu\left( n^\nu e_\nu^{\bar3} e^{\mu\bar3}\right)
+\frac{1}{2}(H-2\dot{\sigma})\delta g^{00}
+a^{-2}e^{4\sigma}\partial_3g_{03}
+\frac{1}{2}a^2e^{-4\sigma}\partial_0\delta g^{33}
+(H-2\dot\sigma)a^2e^{-4\sigma}\delta g^{33}
+\mathcal{O}(\delta^2)\,.
\end{align}
Later we will also use
\begin{align}
\int dtd^3x\,a^3f(t)D_{\mu\nu} e^{\mu\bar3}n^\nu
\label{convenient}
&=\frac{1}{2}\int dtd^3x\,a^3f(t)\left[
(H-2\dot{\sigma})\delta g^{00}
-\left(3H+\frac{\dot{f}}{f}\right)a^2e^{-4\sigma}\delta g^{33}+\mathcal{O}(\delta^2)
\right]\,.
\end{align}

\subsection{Tadpole cancellation}

Finally,
let us solve the background equations of motion.
For this purpose,
we here write down the linear order action.
With the field contents and symmetries being given by \eqref{eq:symms}, we can decompose the effective action schematically as
\begin{equation}
S = S_P + S_L + S_{PL} \, .
\end{equation}
Here, $S_P$ breaks the time-diffeomorphism symmetry only, so that it contains the same operators as single-field inflation. Expanding up to the first-order fluctuations gives
\begin{equation}
S_P 
=
\int d^4x\sqrt{-g}
\left[
\frac{M_{\rm Pl}^2}{2}(\delta g^{00}G_{00}+\delta g^{11}G_{11}+\delta g^{22}G_{22}+\delta g^{33}G_{33})
+(\widetilde{c}-\widetilde\Lambda)-\widetilde{c}\delta g^{00}+\mathcal{O}(\delta^2)
\right]
\,.
\end{equation}
On the other hand, $S_L$ breaks the local Lorentz invariance, in the current case the rotation symmetry. In particular, $S_L$ represents terms which exist even if diffeomorphism symmetries are unbroken. Their lowest few terms in the fluctuations and their derivatives are
\begin{align}
S_{L}&=\int d^4x\sqrt{-g}
\left[
-\frac{\beta_1}{4}\left(\partial_\mu e_\nu^{\bar3}-\partial_\nu e_\mu^{\bar3}\right)^2
-\frac{\beta_2}{2} \left(\nabla^\mu e_\mu^{\bar3}\right)^2
-\frac{\beta_3}{2}\left(e^{\nu\bar3}\nabla_\nu e_\mu^{\bar3}\right)^2
-\frac{\beta_4}{2}\left(\nabla_\mu e_\nu^{\bar3}\right)^2
\right]\,.
\end{align}
It follows from the relation~\eqref{n^me_m}
that the $\beta_2$ coupling starts from the second order in fluctuations.
The $\beta_1$ coupling can be reformulated as
\begin{align}
\nonumber
\left(\partial_\mu e_\nu^{\bar3}-\partial_\nu e_\mu^{\bar3}\right)^2
&=\left(D_{\mu\nu}-D_{\nu\mu}\right)^2
+4(H-2\dot{\sigma})(D_{\mu\nu}-D_{\nu\mu}) e^{\mu\bar3}n^\nu
-2(H-2\dot{\sigma})^2
-2(H-2\dot{\sigma})^2 (n^\mu e_\mu^{\bar3})^2
\\*
&=
4(H-2\dot{\sigma})D_{\mu\nu} e^{\mu\bar3}n^\nu
-2(H-2\dot{\sigma})^2
+\mathcal{O}(\delta^2)
\,,
\end{align}
where we used~\eqref{D_mnn^me^n}
at the second equality.
Using the relation
\begin{align}
e^{\nu\bar3}\nabla_\nu e_\mu^{\bar3}
&=e^{\nu\bar3}D_{\nu\mu}+(H-2\dot{\sigma})n_\mu
\, ,
\end{align}
we also rewrite the $\beta_3$ coupling as
\begin{align}
\nonumber
\left(e^{\nu\bar3}\nabla_\nu e_\mu^{\bar3}\right)^2
&=\left(e^{\nu\bar3}D_{\nu\mu}\right)^2
+2(H-2\dot{\sigma})D_{\mu\nu}e^{\mu\bar3}n^\nu -(H-2\dot{\sigma})^2
\\
&=2(H-2\dot{\sigma})D_{\mu\nu}e^{\mu\bar3}n^\nu -(H-2\dot{\sigma})^2
+\mathcal{O}(\delta^2)\,.
\end{align}
Finally,
the $\beta_4$ coupling takes the form
\begin{align}
\nonumber
\left(\nabla_\mu e_\nu^{\bar3}\right)^2
&=(D_{\mu\nu})^2+2(H-2\dot{\sigma})D_{\mu\nu}e^{\mu\bar3}n^\nu-(H-2\dot{\sigma})^2
\\
&=2(H-2\dot{\sigma})D_{\mu\nu}e^{\mu\bar3}n^\nu-(H-2\dot{\sigma})^2+\mathcal{O}(\delta^2)\,.
\end{align}
In total,
the rotation breaking sector can be written as
\begin{align}
S_L&=\int d^4x\sqrt{-g}
\left[
\frac{\beta_1+\beta_3+\beta_4}{2}(H-2\dot{\sigma})^2
-(\beta_1+\beta_3+\beta_4)(H-2\dot{\sigma})D_{\mu\nu}e^{\mu\bar3}n^\nu
+\mathcal{O}(\delta^2)
\right]\,.
\end{align}
We finally consider the mixing sector:
\begin{align}
S_{PL}&=\int d^4x\sqrt{-g}\,\left[
-\gamma_1(t)\left(e^{0\bar3}\right)^2
+\gamma_2(t)e^{0\bar3}\nabla^\mu e_\mu^{\bar3}
+\gamma_3(t)e^{3\mu}n_\nu\nabla_\mu e^{\nu3}
+\gamma_4(t)\delta g^{00}e^{3\mu}n_\nu\nabla_\mu e^{\nu3}
\right]\,.
\end{align}
Using the formulae in the previous subsection,
it is easy to derive
\begin{align}
S_{PL}&=\int d^4x\sqrt{-g}\,\left[
-\gamma_3(H-2\dot{\sigma})
+\gamma_3D_{\mu\nu}e^{3\mu}n^\nu
-\gamma_4(t)(H-2\dot{\sigma})\delta g^{00}
+\mathcal{O}(\delta^2)
\right]\,.
\end{align}

To summarize so far,
we have obtained
\begin{align}
S&=\int d^4x\sqrt{-g}
\left[
\frac{M_{\rm Pl}^2}{2}R
+(c-\Lambda)-c\delta g^{00}+\lambda D_{\mu\nu}e^{\mu\bar3}n^\nu+\mathcal{O}(\delta^2)
\right]
\end{align}
with $c$, $\Lambda$, and $\lambda$ being
\begin{align}
\Lambda&=\widetilde\Lambda-\frac{\beta_1+\beta_3+\beta_4}{2}(H-2\dot{\sigma})^2
+\gamma_3(H-2\dot{\sigma})+\gamma_4(H-2\dot{\sigma})\,,
\\
c&=\widetilde{c}+\gamma_4(H-2\dot{\sigma})\,,
\\
\lambda&=-(\beta_1+\beta_3+\beta_4)(H-2\dot{\sigma})+\gamma_3\,.
\end{align}
Using the relation~\eqref{convenient},
we further reduce the linear-order action to the form
\begin{align}
S&=\int d^4x\sqrt{-g}
\left\{
\frac{M_{\rm Pl}^2}{2}R
+(c-\Lambda)-\left[ c-\frac{\lambda}{2}(H-2\dot{\sigma}) \right] \delta g^{00}
-\frac{1}{2}\left(
3H\lambda+\dot{\lambda}
\right)
a^2e^{-4\sigma}\delta g^{33}
\right\} \,.
\end{align}
The Einstein equations can be read off as
\begin{align}
M_{\rm Pl}^2G_{00}&=\Lambda+c-\lambda(H-2\dot{\sigma})\,,
\\
M_{\rm Pl}^2G_{11}&=-a^2e^{2\sigma}(\Lambda-c)\,,
\\
M_{\rm Pl}^2G_{33}&=-a^2e^{-4\sigma}\left(\Lambda-c
-3H\lambda-\dot{\lambda}
\right)\,.
\end{align}
More explicitly,
we have
\begin{align}
\Lambda+c-(H-2\dot{\sigma})\lambda
&=M_{\rm Pl}^2(3H^2-3\dot{\sigma}^2)\,,
\\
2c-(H-2\dot{\sigma})\lambda&=
M_{\rm Pl}^2\left(-2\dot{H}+3H\dot{\sigma}+\ddot{\sigma}-6\dot{\sigma}^2\right)\,,
\\
3H\lambda+\dot{\lambda}&=-M_{\rm Pl}^2(9H\dot{\sigma}+3\ddot{\sigma})\,.
\end{align}
Note that FLRW spacetime, i.e. $\sigma=0$, can be realized
if we tune $\lambda$ such that
\begin{align}
\dot{\lambda}=-3H\lambda\,.
\end{align}


\section{Slow-roll and small-anisotropy approximations}
\label{AppB}

Let us define
    \begin{equation}
    -\fr{\dot{H}}{H^2}\coloneqq \ep_H\,,\quad
    \fr{\dot{\si}}{H}\coloneqq \fr{1}{3}I\ep_H\,,
    \end{equation}
and assume $|\ep_H|,|I|\ll 1$.
Provided that $\ep_H$ and $I$ are almost constant, i.e.
    \begin{equation}
    \left|\fr{\dot{\ep}_H}{H\ep_H}\right|\ll 1\,,\quad
    \left|\fr{\dot{I}}{HI}\right|\ll 1\,,\quad
    \end{equation}
one obtains
    \begin{equation}
    H=\fr{1}{\ep_H t}\,,\quad
    \dot{\si}=\fr{I}{3t}\,,
    \end{equation}
and also
    \begin{equation}
    a\propto t^{1/\ep_H}\,,\quad
    e^\si\propto t^{I/3}\,.
    \end{equation}
Thus, the spacetime is described by anisotropic power-law inflation.
Note that the conformal time~$\tau$ is written as
    \begin{equation}
    \tau=\int \fr{dt}{a}=-\fr{1+\ep_H}{aH}\,,
    \end{equation}
where the integration constant has been fixed so that $\tau \to 0$ as $t\to +\infty$.

From the background equations of motion~\eqref{bgEinstein00}--\eqref{bgEOMvec}, we have
    \begin{equation}
    \begin{split}
    &V(\bar{\phi})=3\Mpl^2H^2\,,\quad
    \fr{1}{2}\dot{\bar{\phi}}^2=\ep_H\Mpl^2H^2\,, 
    \\
    &f(\bar{\phi})^2\bar{v}^2=\fr{1}{3}I\ep_H\Mpl^2\,,\quad
    \dot{\bar{v}}=2H\bar{v}\,,
    \end{split} \label{approx}
    \end{equation}
at leading order with respect to $\ep_H$ and $I$.
These relations are used to simplify the coefficients appearing in the quadratic action in \S \ref{ssec:nondynamicalh0}.


\bibliographystyle{mybibstyle}
\bibliography{anisoEFT.bib}

\end{document}